\documentclass[twocolumn]{revtex4}
\usepackage{graphicx}
\usepackage{dcolumn}
\usepackage{longtable}

\begin{document}

\title{Simple, empirical order parameter for a first order quantum phase transition in atomic nuclei}

\author{Dennis Bonatsos$^1$, E.A. McCutchan$^2$, R.F. Casten$^2$ and R.J. Casperson$^{2}$}

\affiliation{$^1$Institute of Nuclear Physics, N.C.S.R.
``Demokritos'', GR-15310 Aghia Paraskevi, Attiki, Greece}

\affiliation{$^2$ Wright Nuclear Structure Laboratory, Yale
University, New Haven, CT 06520, USA}

\begin{abstract}

A simple, empirical signature of a first order phase transition in atomic nuclei is presented, the ratio of the energy of the $6^+$
level of the ground state band to the energy of the first excited $0^+$ state. This ratio provides an effective order parameter
which is not only easy to measure, but also distinguishes between first and second order phase transitions and takes on a special
value in the critical region.  Data in the Nd-Dy region show these characteristics. In addition, a repeating degeneracy between
alternate yrast states and successive excited $0^+$ states is found to correspond closely to the line of a first order phase
transition in the framework of the Interacting Boson Approximation (IBA) model in the large $N$ limit, pointing to a possible
underlying symmetry in the critical region.

\end{abstract}

\maketitle

The study of structural evolution in atomic nuclei has witnessed significant developments in recent years. One of the most important has been the discovery of empirical evidence~\cite{phase1,phase2} for quantum phase transitions (QPT) in the equilibrium shape as a function of nucleon number. This has led to the proposal~\cite{E5,X5} and empirical verification~\cite{134Ba,152Sm} of a new class of models, called critical point symmetries (CPS). These, in turn have spurred an abundance of experimental searches~\cite{nd,mo,gd} for nuclei satisfying the predictions of these CPS as well as investigations into the presence of quasidynamical and partial dynamical symmetries at the critical point~\cite{rowe1,ami}.

In contrast to the usual thermodynamic phase transitions~\cite{landau}, QPT occur at zero temperature as a parameter in the Hamiltonian is varied~\cite{gilmore}. They are attracting much attention in a variety of physical systems, including Josephson-junction arrays and quantum Hall-effect systems~\cite{sondhi}. The transition from Bardeen-Cooper-Schrieffer (BCS) pairing correlations~\cite{bardeen} to Bose-Einstein condensation (BEC) as a result of increasing the strength of the pairing interaction in ultracold alkali atoms has been recently seen experimentally~\cite{greiner,jochim,zwierlein}. In nonrigid polyatomic molecules, a transition from rigidly-linear to rigidly-bent shapes has also been studied~\cite{iachello}. A relationship between QPT and quantum entanglement has also attracted much attention~\cite{oster,gu}. Central to this active field is the search for simple empirical signatures of QPT and the identification of their order. The properties of QPT in atomic nuclei are therefore of quite broad interest, especially since nuclei are finite-body systems in which the number of constituents can be varied experimentally and theoretically.

It is the purpose of this Letter to identify a new, easy-to-measure, observable in one such system, atomic nuclei, that acts as an order parameter for QPT in nuclei and which can distinguish first and second order phase transitions. Further, we will show that a particular value of this observable identifies the entire critical region, even for structures not satisfying a specific CPS. Finally we use this value to point to the possible existence of a heretofore undefined new symmetry at the critical point.

\begin{figure}
\center{{\includegraphics[height=100mm]{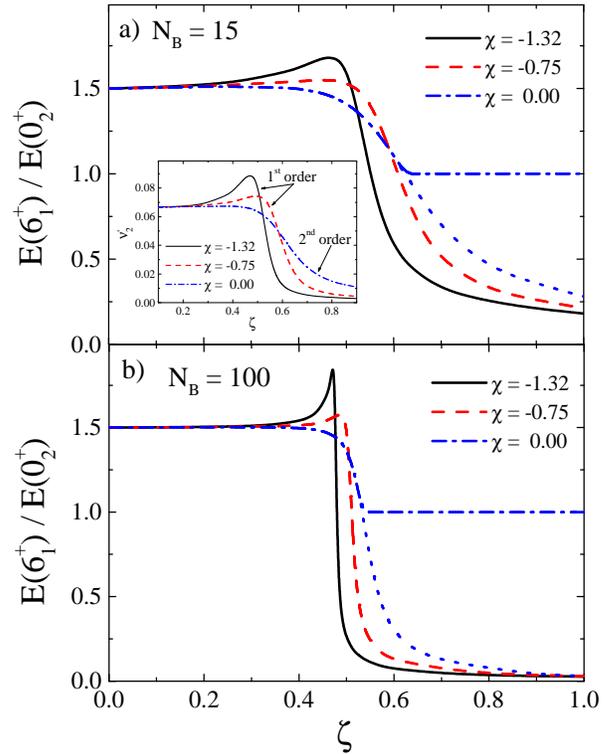}}}
\caption{(Color online) The ratio $E$($6_1^+$)/$E$($0_2^+$) as a
function of $\zeta$ for three values of $\chi$ for (a) $N_B$ = 15
and (b) $N_B$ = 100. The inset to (a) shows the corresponding
behavior for $\nu_2^{\prime}$~\cite{franvic2004}.}
\end{figure}

In the context of an algebraic approach to nuclear structure, the
concept of phase transitions was developed a number of years ago
using the intrinsic state formalism
\cite{gilmore,coherent1,coherent2} of the Interacting Boson
Approximation (IBA) model \cite{iba}. This model, constructed in
terms of the group U(6), has three dynamical symmetries
corresponding to different nuclear shapes:~a spherical nucleus
that can vibrate [U(5)]; an ellipsoidal deformed axially symmetric
rotor [SU(3)], and an axially asymmetric rotor [O(6)]. Nuclei
exist which manifest these symmetries but most nuclei deviate from
them. Spherical to deformed transition regions in the IBA from
U(5) to SU(3) and U(5) to O(6) undergo, in the large boson number
(large valence nucleon number) limit, first and second order phase
transitions, respectively.

More recently, critical point transitions have been described in a geometrical framework, in terms of the Bohr Hamiltonian. These new CPS, E(5) \cite{E5} and X(5) \cite{X5}, correspond to second and first order phase transitions between a vibrator and a rotor, differing in the $\gamma$ degree of freedom. In both cases, an infinite square well potential in the deformation, $\beta$, allows for analytic solutions.

The critical points in the IBA characterize the points in shape transitional regions where various observables \cite{pan03} such as $R_{4/2}$ $\equiv$ $E$($4_1^+$)/$E$($2_1^+$) or electromagnetic transition strengths, $B$($E$2; $2_1^+\rightarrow0_1^+$), $Q$ invariants \cite{volker}, as well as a measure of the wave function entropy \cite{cejnar} vary most rapidly:~their first derivatives have an extremum and their second derivative reverses sign.

With increased interest in QPT in nuclei, several investigations have sought order parameters to distinguish first and second order phase transitions. Two of these are \cite{franvic2004} $\nu_2$, the difference between the expectation values of the number of $d$-bosons in the IBA, $n_d$, in the first excited $0^+$ state and the ground state, and $\nu_2^{\prime}$, proportional to the isomer shift between the first $2^+$ state and the ground state. Both have sharp changes in phase transitional regions and exhibit a different behavior for first and second order phase transitions (see inset to Fig. 1(a)) for small $N_B$.

\begin{figure}
\center{{\includegraphics[height=100mm]{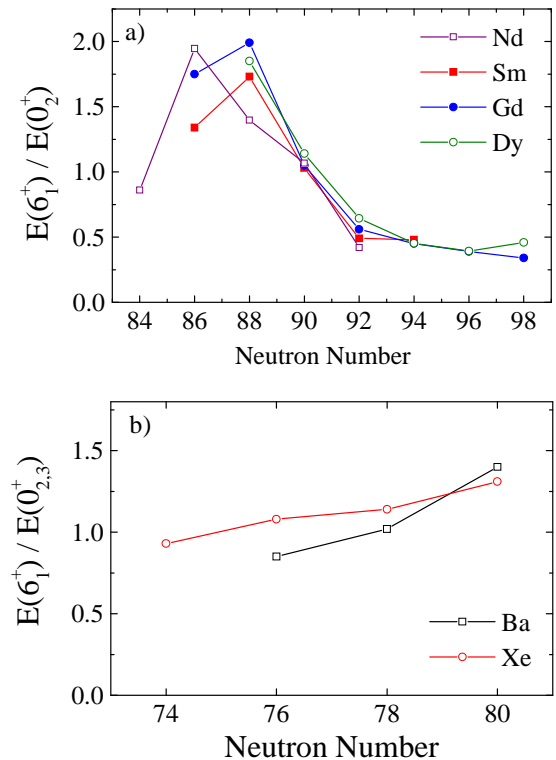}}}
\caption{(Color online)(a) Experimental $E$($6_1^+$)/$E$($0_2^+$)
ratio as a function of neutron number for the Nd, Sm, Gd, and Dy
isotopes. (b) Same for the Xe and Ba isotopes. For smaller neutron
numbers, the $0_3^+$ state was taken in the ratio if its $B$($E$2)
decay was consistent with the $\sigma$ = $N-2$ state. This
corresponds to $N$ = 74 in Xe and $N$ = 76,78 in Ba.}
\end{figure}

However, these differences do not persist in the large $N_B$
thermodynamic limit \cite{pan}. In a more practical vein, Ref.
\cite{rowe} considered the ratio of electromagnetic transition
strengths, $B_{4/2}$ $\equiv$
$B$($E$2;$4_1^+\rightarrow2_1^+$)/$B$($E$2;$2_1^+\rightarrow0_1^+$),
in the IBA which exhibits a peak prior to the critical region for
the U(5)-SU(3) transition, similar to $\nu_2^{\prime}$.  Ref.
\cite{zhang} determined that a difference between U(5)-SU(3) and
U(5)-O(6) transitions persists for $B_{4/2}$ and another $B(E2)$
ratio up to $N_B$ = 50. These signatures however, are often
difficult to measure, and have substantial uncertainties,
particularly far off stability. Moreover, the $B_{4/2}$ ratio
varies only slightly as a function of structure, typically in the
range 1.4-1.7 only.

The present work presents four key results. We identify an order parameter based on the energies of the first excited $0^+$ state and the $6^+$ level of the ground state band. We show that the ratio $E$($6_1^+$)/$E$($0_2^+$) can distinguish between first and second order phase transitions and that this distinction persists in the large $N_B$ limit. We also identify, for the first time, a signature of the entire critical region which closely characterizes the line of the first order phase transition in the IBA in the large $N_B$ limit, that is a signature which identifies a nucleus as lying near the critical point, regardless of the $\gamma$-dependence of the potential.  Finally, we also discuss a possible underlying symmetry in the critical region of the IBA.

Calculations for this study were performed using the extended consistent $Q$~\cite{cqf,ecqf} formalism (ECQF) of the IBA with a Hamiltonian given by \cite{ham1}

\begin{equation}\label{eq:e2}
H(\zeta,\chi) = c \left[ (1-\zeta) \hat n_d -{\zeta\over 4 N_B}
\hat Q^\chi \cdot \hat Q^\chi\right],
\end{equation}

\noindent where $\hat n_d = d^\dagger \cdot \tilde d$, $\hat
Q^\chi = (s^\dagger \tilde d + d^\dagger s) +\chi (d^\dagger
\tilde d)^{(2)},$ $N_B$ is the number of valence bosons, and $c$
is a scaling factor. The above Hamiltonian contains two
parameters, $\zeta$ and $\chi$, with $\zeta$ ranging from 0 to 1,
and $\chi$ ranging from 0 to $-\sqrt{7}/2$. In this
parameterization, the three dynamical symmetries are given by
$\zeta$ = 0, any $\chi$ for U(5), $\zeta$ = 1, $\chi$ =
$-\sqrt{7}/2$ for SU(3) and $\zeta$ = 1.0, $\chi$ = 0.0 for O(6).
Calculations were performed with the code IBAR \cite{ibar} which
allows boson numbers up to 250.

In Fig.~1, the results of calculations for the ratio
$E$($6_1^+$)/$E$($0_2^+$) are given for $N_B$ = 15 and 100. The
calculations include the U(5)-SU(3) transition ($\chi$ =
$-\sqrt{7}/2$), the U(5)-O(6) transition ($\chi$ = 0.0) and an
intermediate $\chi$ value. This ratio is 1.5 in the U(5) limit and
approaches zero in the large $N_B$ limit of SU(3) as the $0_2^+$
state rises in energy. For the case of the U(5)-O(6) transition
($\chi$ = 0), we present two sets of calculations. For $N_B$ = 15,
above $\zeta$ $\sim$ 0.6, the dotted line follows the $0^+$ state
belonging to the $\sigma$ = $N$-2 family of the O(6) symmetry
whereas, the dash-dotted line follows the $0^+$ belonging to the
$\sigma$ = $N$ family (the actual $0_2^+$ level).

For small $N_B$, Fig.~1(a), $E$($6_1^+$)/$E$($0_2^+$) exhibits a
modest peak \textit{before} the first order phase transition point
($\chi$ = $-\sqrt{7}/2$, solid curve) followed by a sharp decrease
across the phase transition. For the second order case, it
maximizes at U(5) and gradually decreases with increasing $\zeta$.
This behavior is identical to that for $\nu_2^{\prime}$ in
Fig.~1(a) [inset]. The intermediate $\chi$ value also exhibits a
small rise before the phase transition. Thus, as $\chi$ approaches
the second order case, the unique features of the first order
phase transition diminish, but are still present, although they
may be difficult to distinguish in finite nuclei.

Figure 1(b) illustrates that the above behavior persists and is enhanced in the large $N_B$ limit. The features of $E$($6_1^+$)/$E$($0_2^+$) are almost identical to the behavior of the $B_{4/2}$ ratio discussed in Ref.~\cite{rowe}. The transition region becomes sharper and occurs for a narrower range of $\zeta$ values, while the peak prior to the phase transition increases in magnitude. With increasing $N_B$, the $E$($6_1^+$)/$E$($0_2^+$) ratio has values close to the $\zeta$ = 0 and $\zeta$ = 1 limiting cases for $\zeta$ values just outside of the transition region.

\begin{figure}
\center{{\includegraphics[height=90mm]{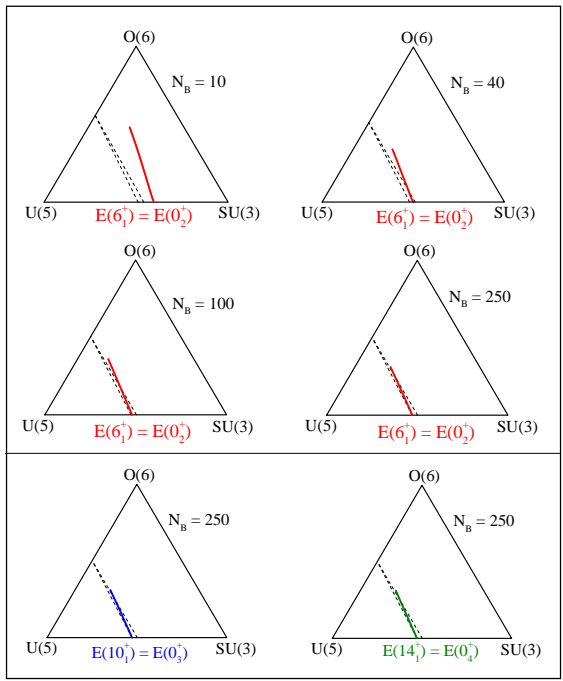}}}
\caption{(Color online) (Top) Line of degeneracy between the
$0_2^+$ and $6_1^+$ levels (solid red) for $N_B$ = 10, 40, 100,
and 250 in the IBA triangle. (Bottom) Line of degeneracy between
the $0_3^+$ and $10_1^+$ levels (solid blue) for $N_B$ = 250
(left) and between the $0_4^+$ and $14_1^+$ levels (solid green
line) for $N_B$ = 250 (right) in the IBA triangle. The dashed
lines denote the critical region in the IBA obtained in the large
$N_B$ limit from the intrinsic state
formalism~\cite{gilmore,coherent1,coherent2}.}
\end{figure}

It is also important to determine if this behavior is observed in
nuclei. In Fig.~2 (a), the $E$($6_1^+$)/$E$($0_2^+$) ratio for the
Nd-Dy isotopes is plotted as a function of neutron number. The Nd,
Sm, and Gd isotopic chains are well established as undergoing a
transition from spherical to deformed shapes, with $^{150}$Nd,
$^{152}$Sm, and $^{154}$Gd lying close to the phase transitional
point \cite{152Sm,nd,gd} and approximately manifesting X(5).
Interestingly, as in Fig.~1, $E$($6_1^+$)/$E$($0_2^+$) peaks just
prior to the phase transitional nuclei, followed by a sharp drop
occurring for the proposed critical point nuclei. (In $^{150}$Nd,
the rise occurs somewhat earlier.) The Dy isotopes behave
similarly to the Sm and Gd isotopes, although data on $N$ $\leq$
86 is lacking. Figure 2 (b) shows the
$E$($6_1^+$)/$E$($0_{2,3}^+$) ratio for the Xe and Ba isotopes
which lie in a $\gamma$-soft region between the U(5) and O(6)
symmetries that contains only a second order phase transition.
Consistent with this, no discontinuity is observed.

\begin{figure}
\center{{\includegraphics[height=45mm]{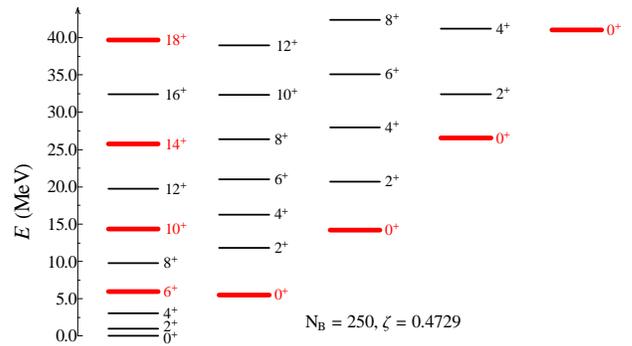}}}
\caption{(Color online) Energies of low-lying states (normalized
to $E$($2_1^+$)=1) of the Hamiltonian of Eq. (1) with
$\chi$=$-\sqrt{7}/2$, $\zeta$=0.4729, and $N_B$=250. The parameter
$\zeta$ was chosen to reproduce the approximate degeneracy of
$E$($0_2^+$) and $E$($6_1^+$).}
\end{figure}

One key signature of the X(5) critical point symmetry is a near
degeneracy of the $6_1^+$ level of the ground state band and the
first excited $0^+$ state. Close inspection of Fig.~1(b) shows
that the $E$($6_1^+$)/$E$($0_2^+$) ratio passes through a value of
1.0 close to the phase transition point. This suggests that a
degeneracy between $E$($6_1^+$) and $E$($0_2^+$) is related to the
critical point. We note that this is reflected empirically in
Fig.~2, where the $N$ = 90 nuclei also have
$E$($6_1^+$)/$E$($0_2^+$) very close to unity. In Fig.~3(top), the
line of $E$($0_2^+$) = $E$($6_1^+$) is given in the IBA triangle
for $N_B$ = 10, 40, 100, and 250. For each boson number, there is
a single trajectory in the triangle which corresponds to this
degeneracy condition. With increasing $N_B$, this line quickly
approaches the phase transition region of the IBA, intersecting it
almost exactly for $N_B$ = 250. Thus, it appears that a degeneracy
of $E$($0_2^+$) and $E$($6_1^+$) is a signature of the $line$ of
first order phase transitions in the IBA in the large $N_B$ limit.

Degeneracies are often associated with underlying symmetries. It
is therefore of considerable interest that the critical point of
the IBA in the large $N_B$ limit exhibits the same degeneracy
between $E$($0_2^+$) and $E$($6_1^+$) as approximately given by
X(5).  To further investigate the presence of underlying
symmetries in the phase transition region of the IBA, Fig. 4 gives
a detailed level scheme from the IBA with $\chi$ = $-\sqrt{7}/2$
and $\zeta$ = 0.4729 (which approximately reproduces the
degeneracy of $E$($0_2^+$) and $E$($6_1^+$)). This point lies very
close to the critical point, $\zeta_{crit}$ = 0.4721, obtained in
the exact infinite $N_B$ limit.  Clearly, additional degeneracies
are also present. Successive ground band members with $J$ $>$ 2
with $J$/2 odd, are nearly degenerate with successive excited
$0^+$ states [$e.g.$, ($6^+_1$, $0_2^+$), ($10^+_1$, $0_3^+$),
($14^+_1$, $0_4^+$)]. Included in Fig.~3 (bottom) are lines of
degeneracy between the $0_3^+$ and $10_1^+$ levels (left) and the
$0_4^+$ and $14_1^+$ levels (right). They too persist along the
entire phase transition region.

The continuation of the degeneracies, that appear at the
intersection of the critical line and the U(5)-SU(3) leg of the
triangle, into the interior of the triangle is reminiscent of the
quasi-dynamical symmetries \cite{quasi1,quasi2} associated with
the U(5)-O(6) and U(5)-SU(3) transitions in the IBA, as well as in
the vibrator to $\gamma$-unstable rotor transition in the Bohr
model~\cite{quasi3}. If the degeneracies in the interior can
indeed be explained by quasi-dynamical symmetries, it would
significantly expand the relevance of such ideas. In the present
case, of course, the underlying symmetry, exact or approximate,
has not yet been identified and needs further investigation.

In conclusion, we have identified a simple, empirical observable,
the ratio $E$($6_1^+$)/$E$($0_2^+$) of the energies of the $6_1^+$
state and the $0_2^+$ state which can serve as an order parameter
identifying phase transitional behavior, and whose behavior can
distinguish between first and second order phase transitions.
Experimental data on the Sm and Gd isotopes, which are well known
to exhibit first order phase transitional behavior, exhibit
exactly the behavior predicted by the IBA for the
$E$($6_1^+$)/$E$($0_2^+$) ratio. Nuclei in second order
transitional regions do not.

Beyond its applicability as an effective order parameter, it was
also found that a particular value of the ratio,
$E$($6_1^+$)/$E$($0_2^+$) = 1.0, provides a signature of the
critical region which persists along the line of a first order
phase transition cutting across the triangle in the IBA in the
large $N_B$ limit. The predicted degeneracy of  $E$($6_1^+$) and
$E$($0_2^+$) in the IBA critical region for large $N_B$ is
consistent with the X(5) critical point symmetry. Finally, the
possibility of an underlying quasi-dynamical symmetry, reflected
in a recurring pattern of degeneracies, in the critical region of
the IBA was discussed.

Useful discussions with F.~Iachello and N.V.~Zamfir are
acknowledged. We are grateful to E. Williams and V. Werner for
discussions of their work on the IBA model for very large boson
numbers, including their role in the development of the code IBAR,
which made the present study possible.  This work was supported by
U.S.~DOE Grant No.~DE-FG02-91ER-40609.



\begin{thebibliography}{99}

\bibitem{phase1} F. Iachello, N.V. Zamfir, and R.F. Casten, Phys.
Rev. Lett. {\bf 81}, 1191 (1999).

\bibitem{phase2} R.F. Casten, Dimitri Kusnezov, and N.V. Zamfir,
Phys. Rev. Lett. {\bf 82}, 5000 (1999).

\bibitem{E5} F. Iachello, Phys. Rev. Lett. {\bf 85}, 3580 (2000).

\bibitem{X5} F. Iachello, Phys. Rev. Lett. {\bf 87}, 052502 (2001).

\bibitem{134Ba} R.F. Casten and N.V. Zamfir, Phys. Rev. Lett. {\bf
85} 3584 (2000).

\bibitem{152Sm}  R.F. Casten and N.V. Zamfir, Phys. Rev. Lett. {\bf
85} 052503 (2001).

\bibitem{nd} R. Kr\"{u}cken {\it et al.,} Phys. Rev. Lett. {\bf
88}, 232501 (2002).

\bibitem{mo} P.G. Bizzeti and A.M. Bizzeti-Sona, Phys. Rev. C {\bf
66}, 031301(R) (2002).

\bibitem{gd} D. Tonev {\it et al.,} Phys. Rev. C. {\bf
69}, 034334 (2004).

\bibitem{rowe1} D.J. Rowe, Phys. Rev. Lett. {\bf 93}, 122502
(2004).

\bibitem{ami} A. Leviatan, Phys. Rev. Lett. {\bf 98}, 242502
(2007).

\bibitem{landau} L.D. Landau and E.M. Lifshitz, {\it Statistical
Physics} (Butterworth-Heinemann, Oxford, England, 1951).

\bibitem{gilmore} D.H. Feng, R. Gilmore, and S.R. Deans, Phys.
Rev. C {\bf 23}, 1254 (1981).

\bibitem{sondhi} S.L. Sondhi, S.M. Girvin, J.P. Carini, and D.
Shahar, Rev. Mod. Phys. {\bf 69}, 315 (1997).

\bibitem{bardeen} J. Bardeen, L.N. Cooper, and J.R. Schrieffer,
Phys. Rev. {\bf 106}, 162 (1957); {\bf 108}, 1175 (1957).

\bibitem{greiner} M. Greiner, C.A. Regal, and D.S. Jin, Nature
(London) {\bf 426}, 537 (2003).

\bibitem{jochim} S. Jochim {\it et al.,} Science {\bf 302}, 2101
(2003).

\bibitem{zwierlein} M.W. Zwierlein {\it et al.,} Phys. Rev. Lett.
{\bf 91}, 250401 (2003).

\bibitem{iachello} F. Iachello, F. P\'{e}rez-Bernal, and P.H.
Vaccaro, Chem. Phys. Lett. {\bf 375}, 309 (2003).

\bibitem{oster} A. Osterloh {\it et al.,} Nature {\bf 416}, 608
(2002).

\bibitem{gu} S.J. Gu, S.S. Deng, Y.Q. Li, and H.Q. Lin, Phys. Rev.
Lett. {\bf 93}, 086402 (2004).

\bibitem{coherent1} J.N. Ginocchio and M.W. Kirson, Phys. Rev. Lett. {\bf 44}, 1744 (1980).

\bibitem{coherent2} A.E.L. Dieperink, O. Scholten, and F. Iachello, Phys. Rev. Lett. {\bf 44}, 1747 (1980).

\bibitem{iba} F. Iachello and A. Arima, {\it The Interacting Boson
Model} (Cambridge University, Cambridge, England, 1987).


\bibitem{pan03} Feng Pan, J.P. Draayer, and Yanan Luo, Phys. Lett.
B {\bf 576}, 297 (2003).

\bibitem{volker} V. Werner, P. von Brentano, R.F. Casten, and J.
Jolie, Phys. Lett. B {\bf 527}, 55 (2002).

\bibitem{cejnar} P. Cejnar and J. Jolie, Phys. Rev. E {\bf 58},
387 (1998).


\bibitem{franvic2004} F. Iachello and N.V. Zamfir, Phys. Rev.
Lett. {\bf 92}, 212501 (2004).

\bibitem{pan} Feng Pan, Yu Zhang, and J.P. Draayer, J. Phys. G:
Nucl. Part. Phys. {\bf 31}, 1039 (2005).

\bibitem{rowe} D.J. Rowe, P.S. Turner, and G. Rosensteel, Phys.
Rev. Lett. {\bf 93}, 232502 (2004).

\bibitem{zhang} Yu Zhang, Zhan-feng Hou, and Yu-xin Liu, Phys.
Rev. C {\bf 76}, 011305(R) (2007).

\bibitem{cqf} D.D. Warner and R.F. Casten, Phys. Rev. Lett. {\bf
48}, 1385 (1985).

\bibitem{ecqf} P.O. Lipas, P. Toivonen, and D.D. Warner, Phys.
Lett. B {\bf 155}, 295 (1985).

\bibitem{ham1}  V. Werner {\it et al.,} Phys. Rev. C {\bf 61}, 021301(R) (2000).


\bibitem{ibar} R.J. Casperson, IBAR code (unpublished).

\bibitem{quasi1} D.J. Rowe, Nucl. Phys. A {\bf 745}, 47 (2004).

\bibitem{quasi2} G. Rosensteel and D.J. Rowe, Nucl. Phys. A {\bf
759}, 92 (2005).

\bibitem{quasi3} P.S. Turner and D.J. Rowe, Nucl. Phys. A {\bf
756}, 333 (2005).



\end{thebibliography}
\end{document}